\documentstyle[aps,pre,multicol,eqsecnum]{revtex}

\begin{document}
\draft
\title{Viscous Instanton for Burgers' Turbulence}
\author{E. Balkovsky$^a$, G. Falkovich$^a$, I. Kolokolov$^b$ 
and V. Lebedev$^{a,c}$}
\address{$^a$ Physics Department, Weizmann Institute of Science,
Rehovot 76100, Israel \\ $^b$ Budker Institute of
Nuclear Physics, Novosibirsk 630090, Russia
\\ $^{c}$ Landau Inst. for Theor. Physics, Moscow, Kosygina 2,
117940, Russia}
\date{\today}
\maketitle

\begin{abstract}

We consider 
the tails of probability density functions (PDF) for different characteristics 
of velocity that satisfies Burgers equation driven by a large-scale force.
The saddle-point approximation is employed in the path integral
so that the calculation of the PDF tails
boils down to finding the special field-force configuration (instanton) that 
realizes the extremum of probability. We calculate high moments of the velocity 
gradient $\partial_xu$ and find out that they correspond to the PDF with 
$\ln[{\cal P}(\partial_xu)]\propto-(-\partial_xu/{\rm Re})^{3/2}$ where 
${\rm Re}$ is the Reynolds number. That stretched exponential form is valid 
for negative $\partial_xu$ with the modulus much larger than its 
root-mean-square (rms) value. The respective tail of PDF for negative velocity 
differences $w$ is steeper than Gaussian, 
$\ln{\cal P}(w)\sim-(w/u_{\rm rms})^3$, as well as single-point
velocity PDF  $\ln{\cal P}(u)\sim-(|u|/u_{\rm rms})^3$.
For high velocity derivatives $u^{(k)}=\partial_x^ku$, the general
formula is found: $\ln{\cal P}(|u^{(k)}|)\propto
-(|u^{(k)}|/{\rm Re}^k)^{3/(k+1)}$. 

\end{abstract}

\section{Introduction}

The forced Burgers equation
\begin{equation}
\partial_t u+u\partial_x u 
-\nu\partial_x^2 u=\phi
\label{be1} \end{equation}
describes the evolution of weak one-dimensional acoustic perturbations 
in the reference frame moving with the sound velocity \cite{Burg}. It is 
natural to assume the external force $\phi$ to be $\delta$-correlated 
in time in that frame:
\begin{equation}
\langle\phi(t_1,x_1)\phi(t_2,x_2)\rangle=
\delta(t_1-t_2)\chi(x_1-x_2).
\label{be2} \end{equation}
Then the statistics of $\phi$ is Gaussian and is entirely
characterized by (\ref{be2}). We are interested in turbulence excited by
a large-scale pumping with the typical correlation length $L$ so that $\chi$
does not essentially change at $x\lesssim L$ and goes to zero where $x>L$. 
Besides $L$, the correlation function $\chi$ is characterized by the parameter 
$\omega=[-(1/2)\chi''(0)]^{1/3}$ having the dimensionality of frequency. 
Then e.g. $\chi(0)\sim L^2\omega^3$. Developed turbulence corresponds 
to the large value of Reynolds number ${\rm Re}=L^2\omega/\nu$.

The physical picture of Burgers turbulence is quite clear: arbitrary localized 
perturbation evolves into shock wave with the viscous width of the front,
that gives $k^{-2}$ for the energy spectrum at ${\rm Re}\gg kL\gg1$ 
\cite{Burg,Kra68}. The presence of shocks leads to a strong 
intermittency, PDF of velocity gradients is substantially non-Gaussian 
\cite{GK93} and there is an extreme anomalous scaling for the moments of 
velocity differences $w=u(\rho)-u(-\rho)$:
$\langle w^n\rangle\propto (\rho/L)$ for $n>1$ \cite{Goto94}. 
Simplicity of the equation and transparency of underlying physics make it 
reasonable to hope that consistent formalism for the description of 
intermittency could be developed starting from Burgers equation
\cite{Pol95,BMP95,CY95} as well as from the problem of passive scalar 
advection \cite{95KYC,95CFKL,95GK,95SS}. That would be an appropriate 
way to mark centenary of Burgers' birth.

The most striking manifestation of intermittency is the strong deviation of 
the PDF tails from Gaussian or the behavior of high moments. The PDF's
${\cal P}(\partial_xu)$ and ${\cal P}(w)$ are not symmetric, the asymmetry 
is due to the simple fact that positive gradients are smeared while the 
steepening of negative gradients could be stopped only by viscosity.
It has been realized recently  that those tails can be found by 
considering the saddle-point configurations (we call them instantons) in the 
path integral determining the PDF \cite{95FKLM}. The instanton formalism has 
been applied to the Burgers equation by Gurarie and Migdal \cite{95GM} who 
found the right tail $\ln{\cal P}(w)\sim-[w/(\omega\rho)]^3$ determined by 
practically inviscid behavior of smooth ramps between the shocks. That cubic 
right tail has been first predicted by Polyakov from the conjecture on the 
operator product expansion \cite{Pol95}, it corresponds to the same right tail
for the gradients $\ln{\cal P}(\partial_xu)\sim -(\partial_xu/\omega)^3$ which
also has been derived from the mapping closure by Gotoh and Kraichnan 
\cite{GK96}. Here we find the viscous instantons that give the left tails of 
PDF's and single-point velocity PDF.

Even though some calculations are lengthy, the simple picture appears as a 
result. Since white forcing pumps velocity by the law $w^2\propto \phi^2t$ 
while the typical time of growth is restricted by the breaking time 
$t\sim L/w$ then the Gaussianity of the forcing 
$\ln{\cal P}(\phi)\propto-\phi^2/\chi(0)$ leads to 
$\ln{\cal P}(w)\sim-[|w|/(L\omega)]^3$. At a shock, 
$w^2\simeq-\nu \partial_xu$ so that 
$\ln{\cal P}(\partial_xu)\sim-[-\partial_xu/(\omega{\rm Re})]^{3/2}$.
These simple estimates are confirmed below by consistent calculations.

\section{Saddle-Point Approximation}

A general instanton formalism for the description of 
the high-order correlation functions and of the tails of probability 
density functions has been developed in \cite{95FKLM}. Here, we will be 
interested in the high-order moments of the velocity gradient 
$\partial_xu$, difference $w=u(x=\rho)-u(x=-\rho)$ and the velocity itself:
\begin{eqnarray} &&
\langle[\partial_xu]^n\rangle={\cal Z}_0^{-1}
\int {\cal D}u{\cal D}p\exp\left\{i{\cal I}+
n\ln[\partial_xu(0,0)]\right\}\,,
\label{si1} \\ &&
\langle w^n\rangle={\cal Z}_0^{-1}
\int {\cal D}u{\cal D}p\exp\left\{i{\cal I}+
n\ln[u(0,\rho)-u(0,-\rho)]\right\}\,,
\label{sio}\\ &&
\langle u^n\rangle={\cal Z}_0^{-1}
\int {\cal D}u{\cal D}p\exp\left\{i{\cal I}+n\ln u(0,0)\right\}\,.
\label{siv} \end{eqnarray}  
Here, ${\cal Z}_0$ is the normalization constant and the effective
action ${\cal I}=\int dt\,{\cal L}(u,p)$. The form of the Lagrangian ${\cal L}$ 
for (\ref{be1},\ref{be2}) is determined by the dynamical equations 
\cite{73MSR,76Dom,76Jan}:
\begin{eqnarray} &&
{\cal L}=\int dx\,(p\partial_t u+pu\partial_xu-\nu p\partial_x^2u)
+\frac{i}{2}\int dx_1dx_2p_1\chi_{12}p_2\ .
\label{va1} \end{eqnarray}
The expressions (\ref{si1},\ref{sio}) with (\ref{va1}) are formally exact 
relations.

\subsection{Saddle-Point Equations}

The main idea implemented here is that the high-order correlation functions 
(for $n\gg1$) are determined by the saddle-point configurations of the 
path integrals (\ref{si1},\ref{sio}). The corresponding saddle-point
equations are
\begin{eqnarray} &&
\partial_t u+u\partial_xu-\nu\partial_x^2u
=-i\int dx'\,\chi(x-x')p(x'),
\label{va2} \\ &&
\partial_t p+u\partial_xp
+\nu\partial_x^2p
=\delta(t)\lambda(x).
\label{vam} \end{eqnarray} 
We should substitute here
\begin{eqnarray} &&
\lambda\Rightarrow[{in}/{a(0)}]\delta'(x)
\quad a(0)=\partial_xu(0,0) \quad {\rm for\ gradients},
\label{sh1} \\ &&
\lambda\Rightarrow-({in}/{w})[\delta(x-\rho)-\delta(x+\rho)]
\quad {\rm for\ differences}.
\label{sh2} \\ &&
\lambda\Rightarrow-[{in}/{u}(0,0)]\delta(x)
\quad {\rm for\ velocity}. 
\label{sh3}\end{eqnarray}
A solution of the equations (\ref{va2},\ref{vam}) determines the
saddle-point value of the arguments of the exponent in (\ref{si1},\ref{sio}) 
and consequently the corresponding correlation functions in the
steepest descend approximation.

The equation (\ref{va2}) can be solved with arbitrary
initial condition at remote negative time. The condition
for the equation (\ref{vam}) should be formulated for the final 
time \cite{95FKLM}. It is $p=0$ as it follows from the variation 
of the action (\ref{va1}) over $u$ at the final time, since
the corresponding contribution to the action originating from
$\int dt\,dx\, p\partial_t u$ is $\int dx\, p\delta u$.
That means that $p=0$ at $t>0$. 
Thus we can solve the system of equations (\ref{va2}) and
\begin{equation}
\partial_t p+u\partial_xp
+\nu\partial_x^2p=0,
\label{va3} \end{equation}
at negative time $t$ only. The role of the last term in (\ref{vam}) is then
reduced to the initial (or more precisely final) condition $p(x)=-\lambda(x)$ 
imposed on $p$ at $t=-0$. The initial condition imposed on $u$ at a remote
negative time is forgotten after a finite time and does not influence
the answer. From the other hand the final condition on the field $p$ 
is also forgotten at a finite negative time if to move backwards in time. 
Thus our solution is characterized by a long stationary stage where both
fields $u$ and $p$ are zero. We will refer to that stage as 
vacuum since it is realized at the absence of the source $\lambda$ 
in the right-hand side of (\ref{vam}). In the following we will treat the
vacuum value $u=0$ as the initial condition for the field $u$. As to the
field $p$ it grows from a small perturbation at moving in the positive 
direction in time due to the negative sign at the viscous term in (\ref{va3}).
We can say that our instanton solution describes the instability inherent to
the vacuum.

The equations (\ref{va2},\ref{va3}) are invariant under 
$x\to-x$, $u\to-u$, $p\to-p$. Thus for gradients and differences,  
the antisymmetry of the boundary term $\lambda$ leads to 
an antisymmetric solution $u(x)=-u(-x)$, $p(x)=-p(-x)$.
For velocity, the solution will be asymmetric.

In the saddle-point approximation
\begin{equation}
\langle[\partial_xu]^n\rangle
\sim\exp(i{\cal I}_{\rm extr})[a(0)]^n, \quad 
\langle w^n\rangle\sim\exp(i{\cal I}_{\rm extr})
[w(0)]^n \quad ,
\langle u^n\rangle\sim\exp(i{\cal I}_{\rm extr}) [u(0,0)]^n,
\label{vaa} \end{equation}
where we should substitute the corresponding values found
as a solution of (\ref{va2},\ref{vam}) and the saddle-point
value of the effective action.
Of course, besides the saddle-point contributions, there exist fluctuation
corrections to the path integrals (\ref{si1}--\ref{siv}). The
corrections at the vacuum stage are compensated by the normalization
constant ${\cal Z}_0$ since the vacuum just corresponds to $n=0$ that is 
to $\lambda=0$. The contribution of the fluctuations around the instanton 
(associated with nonzero saddle-point values of the fields $u$ and $p$) 
is finite and small in comparison with the saddle-point one if $n\gg1$.

\subsection{Conservation Laws}

The equations (\ref{va2},\ref{va3}) are Lagrangian equations and the 
Lagrangian (\ref{va1}) does not explicitly depend on time.
Then the ``energy'' $E$ is conserved which can be found by
the standard canonical transformation:
\begin{equation}
iE=\int dt\,p\partial_tu-{\cal L}.
\label{vae3} \end{equation}
The explicit expression for $E$ can be obtained from (\ref{va1}):
\begin{equation}
E=i\int dx\, (pu\partial_x u+\nu\partial_x p\partial_x u)
-(1/2)\int dx_1\,dx_2\,p_1\chi_{12}p_2\ .
\label{vae} \end{equation}
Since the Lagrangian (\ref{va1}) does not explicitly depend on
coordinates then the ``momentum'' $J$ is conserved as well:
\begin{equation}
iJ=\int dx\,p\partial_xu.
\label{va9}  \end{equation}
Because of the conservation laws we should treat 
solutions of (\ref{va2},\ref{vam}) with zero values of both $E$ and $J$ since 
they are zero in the vacuum. Using $E=0$ we conclude from 
(\ref{va1},\ref{vae},\ref{vae3}) that the saddle-point value of the effective 
action figuring in (\ref{vaa}) is
\begin{equation} 
{\cal I}_{\rm extr}=\int dt\,dx\,p\partial_tu.
\label{vae2} \end{equation}

Any conservation law is an important heuristic tool to be used at the 
beginning to understand general properties of the solutions. Let us consider 
the final instant $t=-0$. Here $p(x)=-\lambda(x)$. Now, substituting that 
into the energy, we can analyze the balance of different terms.
For the gradient, $E=-na(0)+\omega^3n^2/a^2(0)-n\nu \partial_x^3u(0,0)/a(0)=0$.
Striking difference between the cases $a(0)>0$ and $a(0)<0$ is now clearly 
seen from the energy conservation. Indeed, for the case of positive gradient, 
the viscous contribution to the energy is unessential and 
two first terms can compensate each other. Then
one gets $a(0)=\omega n^{1/3}$, which corresponds to the right PDF
tail $\ln{\cal P}(\partial_xu)\sim-(\partial_xu/\omega)^3$ given by an inviscid
instanton (see the next subsection). The instanton that gives $a(0)<0$
cannot exist without viscosity as it is seen from energy conservation.
The same consideration can be developed for velocity differences with the 
same result that positive $w$ comes from an inviscid instanton while negative 
from a viscous one. For the velocity, we use both conservation
laws. Requiring $J$ to be zero we obtain 
$\partial_x u(0,0)=0$. That leads to the following energy at $t=-0$:
$E=\chi(0)n^2/u^2(0)-\nu \partial_x^2u(0,0)$. Without viscous term, energy
conservation cannot be satisfied. Let us emphasize that the answer we shall
obtain for the tails of velocity PDF (as well as that for velocity 
differences) does not contain viscosity while it's consistent derivation 
requires the account of the viscous terms in the equations.

\subsection{Right Tail}

Let us first describe the instantons producing the right tails of the PDFs 
for gradients and differences \cite{Pol95,95GM,GK96}. 
The tails correspond to the positive slope of the
velocity $u$ near the origin. Then one can expect that
the field $u$ is smooth enough and the viscosity plays a minor role
in its dynamics. Due to the initial condition $p=-\lambda$ at $t=0$ the 
field $p$ is localized near the origin. At moving backward in time the 
viscosity will spread the field $p$. Nevertheless,  if $u>0$ at $x>0$ and 
$u<0$ at $x<0$ then the sweeping will tend to ``compress'' the field $p$.
In this situation one can expect that the width of $p$ remains
much smaller than $L$.  Then, it is possible to formulate the closed 
system of equations for the quantities $a(t)=\partial_xu(t,0)$ and
\begin{equation}
c(t)=-i \int dx\,x\,p(t,x),
\label{vm5} \end{equation}
regarding that the width of $p$ is so small that 
the velocity $u$ can be replaced by it's linear expansion term $ax$.
The equation for $c$ can be obtained from (\ref{va3}) if to
multiply it by $x$ and integrate over $x$. The equation for $a$ can
be obtained from (\ref{va2}) where for narrow $p$ we can put 
$\int dx'\chi(x-x')p(t,x')\to-i\partial_x\chi(x)c(t)$ and for small 
$x$ we can substitute $\partial_x\chi(x)\approx-2\omega^3x$. We thus find
\begin{equation}
\partial_tc=2ac, \quad
\partial_ta=-a^2+2\omega^3c.
\label{vca} \end{equation}
The energy conserved by (\ref{vca}) can be
derived from (\ref{vae}) at the same approximation: $E=-ca^2+\omega^3c^2$.
The condition $E=0$ means that the instanton is a separatrix 
solution of (\ref{vca}). 

For gradients $p(-0,x)=-[{in}/{a(0)}]\delta'(x)$ i.e.
$a(0)c(0)=n$. From the energy conservation $\omega^3c^2=ca^2$, 
one gets $a(0)=\omega^3c^2(0)/n=\omega n^{1/3}$. 
The same solution describes also the instanton for differences 
since in this case $p(-0,x)$ is determined by (\ref{sh2}) what leads to 
$w=2a(0)\rho$ and then to the same initial condition $a(0)c(0)=n$. 
Note that the life time of the instanton is $T\sim n^{-1/3}\omega^{-1}$. 

The main term in (\ref{vaa}) is $[a(0)]^n$, that immediately leads to
$\langle(\partial_xu)^n\rangle\sim[a(0)]^n \sim \omega^n n^{n/3}$ 
which gives the right cubic tail of the PDF 
$\ln{\cal P}(\partial_xu)\sim-(\partial_xu/\omega)^3$.
Analogously one can obtain $\ln{\cal P}(w)\sim-[w/(\rho\omega)]^3$ 
found previously in \cite{Pol95,95GM,GK96}. Then one can check that the 
extremum value (\ref{vae2}) is negligible in comparison with $n\ln[a(0)]$:
$$ {\cal I}_{\rm extr}=i\int dt\, c\partial_t a\sim a(0)c(0)=n. $$ 

One can incorporate viscosity and show that it has negligible influence on the 
answer and that the width of $p$ is much less than $L$ through the relevant 
time of evolution giving the main contribution into the action \cite{95GM}. 
The right tails of ${\cal P}(\partial_xu)$ and ${\cal P}(w)$ are 
thus universal i.e. independent of the large-scale properties of the pumping. 

\subsection{Left Tail}
\label{subsec:LT}

The main subject of this paper is the analysis of the instantons that give
the tails of ${\cal P}(u)$ and the left tails of ${\cal P}(\partial_xu)$ 
and ${\cal P}(w)$ corresponding to negative values of $a(0)$, $w$. 
Although the field $p$ is narrow at $t=0$, we cannot use the simple 
system of equations (\ref{vca}) to describe this instanton as
it has been seen from the conservation laws. 
In terms of the function $u$, the reason is in the shock forming near the
origin which cannot be described in terms of the inviscid 
equations. In addition, sweeping by negative velocity slope provides for 
stretching (rather than compression) of the field $p$ at moving backwards 
in time. Thus we should use the system of full differential equations
(\ref{va2},\ref{vam}) to describe the instantons which can be called
viscous instantons. Before presenting analytical consideration, let us 
qualitatively describe the instantons and make simple estimates of 
their parameters. 

We shall see that, apart from a narrow front near $x=0$, the velocity field 
has $L$ as the only characteristic scale of change. The life time $|T|$ of
the instanton is then determined by the moment when
the width of $p$ reaches $L$ due to sweeping by the velocity 
$u_0$: $|T|\sim L/u_0$. Such a velocity $u_0$ itself has been created during 
the life time $|T|$ by the forcing so that $u_0\sim|c|_{max}TL\omega^3$. 
To estimate the maximal value of $|c(t)|$, let us consider the backward 
evolution from $t=0$. We first notice that the width of $p$ (which was zero 
at $t=0$) is getting larger than the width of the velocity front $\simeq u_0/a$
already after the short time $\simeq a^{-1}$. After that time, the 
values of $c$ and $a$ are of order of their values at $t=0$. 
Then, one may consider that $p(t,x)$ propagates (backwards in time) 
in the almost homogeneous velocity field $u_0$ so that 
$\partial_t c=-i\int dx\, xup_x\sim u_0\int dx\, p$. The integral
$\int dx\, p$ can be estimated by it's value at $t=0$ which 
is $n/u_0$. Therefore, we get $c_{max}\simeq nT$ so that
$|T|\simeq n^{-1/3}\omega^{-1}$ and $u_0\simeq L\omega n^{1/3}$. 

Since at the viscosity-balanced shock, the velocity $u_0$ and the gradient $a$ 
are related by $u_0^2\simeq\nu a$ then $a(0)\simeq \omega{\rm Re}\,n^{2/3}$.
The main contribution to the saddle-point value (\ref{vaa})
is again related to the term $[\partial_xu(0,0)]^n$ 
and we find $\langle(\partial_xu)^n\rangle\simeq[a(0)]^n\simeq
(\omega{\rm Re})^n n^{2n/3}$, which corresponds to the following left 
tail of PDF at $\partial_xu\gg \omega{\rm Re}$
\begin{equation} 
{\cal P}(\partial_xu)\propto 
\exp[-C(-\partial_xu/\omega{\rm Re})^{3/2}]\ .
\label{an2} \end{equation}
The velocity difference $w(\rho)$ is $u_0$ for $L\gg\rho\gg x_0$ where the 
width of the shock $x_0\simeq u_0/a(0)\simeq n^{-1/3}{\rm Re}^{-1}$. 
We thus have $\langle w^n\rangle\simeq (L\omega)^nn^{n/3}$
which corresponds to the cubic left tail 
$$ {\cal P}(w)\propto \exp\{-B[w/(L\omega)]^3\}\ .$$
The product $L\omega$ plays the role of $u_{\rm rms}$. 
The instanton describing the velocity statistics is asymmetric since the
velocity maximum should reach $x=0$ at $t=0$ as it has been seen from the
momentum conservation. Nevertheless,
all above estimates concerning $w$ are valid also for $u$ and we obtain
$$ {\cal P}(u)\propto \exp\{-D[u/(L\omega)]^3\}\ .$$ 
The numerical factors $C$, $B$ and $D$ are of order unity, they
are determined by the evolution at $t\simeq T$ i.e. by the behavior of pumping
correlation function $\chi(x)$ at $x\simeq L$. The left tails are 
thus nonuniversal; in addition to $L$ and $\omega$, they are 
determined by the large-scale properties of the pumping.

\section{Analytic Description of the Viscous Instanton}

The formation of the shock near the origin occurs at time $|t|\lesssim |T|$
and at small $x$. At moderate $x$ we can use the inviscid version of 
the equations (\ref{va2},\ref{va3})
\begin{eqnarray} &&
\partial_t u+u\partial_xu
=-i\int dx'\,\chi(x-x')p(x'),
\label{vm1} \\ &&
\partial_t p+u\partial_xp=0\ .
\label{vm2} \end{eqnarray}
That system obeys a rescaling symmetry which
admits the following form of solutions
\begin{equation}
u=Nf_1(t/T,x/L), \quad
p=N^2f_2(t/T,x/L), \quad
T=-N^{-1}\omega^{-1}, 
\label{vm3} \end{equation}
where $f_1,f_2$ satisfy corresponding equations. We will demonstrate
that a unique family with $N$ related to $n$ is realized for our
viscous instanton. To prove that, we should find a solution of the viscous 
equations (\ref{va2},\ref{va3}) at $x\ll L$, $|t|\ll|T|$ and to match 
the solution with (\ref{vm3}) at intermediate scales and time. 

We shall construct this short-scale solution at this section.
Here, we want to note only that the shock formed at small time near the
origin can be in the main approximation described by the stationary
equation $u\partial_xu-\nu\partial_x^2u=0$ which can be derived from
(\ref{va2}) where the time derivative and the pumping term
can be neglected in comparison with large space derivatives. 
Then we find the well known expression
\begin{equation}
u=\frac{2\nu}{x_0}\tanh(x/x_0),
\label{vm4} \end{equation}
describing the field $u$ at $t=0$ and small enough $x$,
the width of the shock $x_0$ should be much less than $L$.
At scales $x\gg x_0$ we return to the behavior (\ref{vm3}) what
gives the estimate $L\omega N$ for the amplitude of (\ref{vm4}). 
Then we find from (\ref{vm4})
\begin{equation}
x_0\sim L {\rm Re}^{-1}N^{-1}, \quad
a(0)\sim \omega N^2{\rm Re}, \quad
w\sim L\omega N,
\label{vm6} \end{equation}
where $w=u(0,\rho)-u(0,-\rho)$ for $x_0\ll\rho\ll L$.

Since the instanton describing the velocity statistics is asymmetric
the shock is not formed near the origin. It can be described by 
the expression $u=(2\nu/x_0)\tanh[(x-\rho)/x_0]+{\rm const}$ 
obtaining from (\ref{vm4}) by shifting. Thus the general conclusions
made above are valid for the instanton but the velocity statistics
needs a separate consideration. The next subsections are devoted to
the instantons describing the statistics of the derivatives and of
the differences and the last subsection presents the instanton for
the velocity statistics.

\subsection{Cole-Hopf Representation}

It is known that the equation (\ref{be1}) can be rewritten
in the linear form by using Cole-Hopf substitution \cite{Burg}. 
To proceed further we shall introduce the substitution for our instanton. 
Namely we reformulate the problem for the new variables $\Psi$ and $P$:
\begin{equation}
\partial_x\Psi=-\frac{u}{2\nu}\Psi, \quad
2\nu\partial_xp=P\Psi.
\label{ha1} \end{equation}
The action (\ref{va1}) is rewritten as
\begin{eqnarray} &&
{\cal I}=\int dt\,dx\, P(\partial_t\Psi-\nu\partial_x^2\Psi)
+\frac{i}{2}\int dt\,dx_1\,dx_2\,\chi_{12}p_1p_2.
\label{ha2} \end{eqnarray}
The energy (\ref{vae}) is now
\begin{eqnarray} &&
E=i\int dx\, \nu \partial_x P\partial_x\Psi
+\frac{i\nu}{2}\int dx\,FP\Psi
\nonumber \\ &&
=-i\int dx\,P\partial_t\Psi
-\frac{i\nu}{2}\int dx\,FP\Psi\,,\mbox{ where}
\label{ha6} \\ && 
\partial_xF=-\frac{i}{2\nu^2}
\int dx'\chi(x-x')p(x').
\label{ha4} \end{eqnarray}
We will fix $F$ by the condition $F(x=0)=0$. Note that for the 
antisymmetric solution (describing the gradients and the differences)
$$ F(x=+\infty)-F(x=-\infty)=-\frac{i}{2\nu^2}
\int dx\,\chi(x)\int dx'\,p(x')=0. $$

The saddle-point equations for the functions $\Psi$ and $P$ read
\begin{eqnarray} &&
\partial_t\Psi-\nu\partial_x^2\Psi+\nu F\Psi=0,
\label{ha7} \\ &&
\partial_t P+\nu\partial_x^2P-\nu FP
-{2\nu}\lambda'(x)\delta(t)\Psi^{-1}=0.
\label{ha3} \end{eqnarray}
Here, we have introduced the term from (\ref{vam}), producing the final 
condition for $P$, into the corresponding equation.
One concludes from (\ref{ha7},\ref{ha3}) that 
\begin{equation}
\partial_t(P\Psi)+\nu\partial_x
(\partial_xP\Psi-P\partial_x\Psi)={2\nu }\delta(t)\lambda'(x),
\label{ho1} \end{equation} 
and the quantity $\int dx\, P\Psi$ is conserved. Actually, the
integral is equal to zero since $P(x)=0$ at $t=+0$. That means that we
can treat functions $p(x)$ tending to zero at $x\to\pm\infty$
since $\int dx\, P\Psi\propto\int dx\,\partial_x p$ in accordance
with (\ref{ha1}). Note that the saddle-point value of the action 
(\ref{vae2}) can be rewritten also in terms of the fields $\Psi$ and $P$:
\begin{equation}
{\cal I}_{\rm extr}=\int dt\,dx\,P\partial_t\Psi.
\label{vae4} \end{equation} 

As we already pointed out,  the spatial width of $p(t,x)$ is small at 
small $|t|$. It enables one to go further in formulating the equations for the
fields $P$ and $\Psi$. That allows us to expand $\chi(x-x')$ in (\ref{ha4}) 
in the series over $x'$. The leading term gives 
\begin{equation}
F=\frac{c}{2\nu^2}[\chi(0)-\chi(x)].
\label{hoa} \end{equation}
Here $c$ is defined by (\ref{vm5}), $c$ can be also
rewritten in terms of the fields $\Psi$ and $P$ 
\begin{equation}
c=\frac{i}{4\nu}\int dx\, x^2 P \Psi .
\label{ho2} \end{equation} 
For $x\ll L$ (\ref{hoa}) is reduced to $F(x)=[\omega^3/(2\nu^2)]cx^2$.
Then, the system (\ref{ha7},\ref{ha3}) can be rewritten as follows
\begin{eqnarray} &&
\partial_t\Psi-\nu\partial_x^2\Psi+\frac{\omega^3}{2\nu}cx^2\Psi=0,
\label{ha8} \\ &&
\partial_t P+\nu\partial_x^2P-\frac{\omega^3}{2\nu}cx^2 P
-{2\nu}\delta(t)\lambda'(x)\Psi^{-1}=0,
\label{ha9} \end{eqnarray}
and the energy (\ref{ha6}) is
\begin{equation}
E=-i\int dx\,P\partial_t\Psi-\omega^3c^2.
\label{hae} \end{equation}

\subsection{Generating Function}

The temporal evolution of $p$ for time where it's width is less than 
$L$ can be effectively described in terms of
the generating function for the moments
\begin{equation}
Y(\eta)=-i\int dx\,\exp(-\eta x^2/2)\partial_x p\ .
\label{ap1} \end{equation}
Note that $Y(0)=0$ and $\partial_\eta Y(0)=c$ where $c$ is determined by
the expression (\ref{vm5}). We shall express the behavior of the moments of 
$p$ via some derivatives of $u$ taken at the initial 
time $t=0$. Since the field $p$ is narrow enough at small time we can
use (\ref{ha8},\ref{ha9}) at $x\ll L$ to calculate the
generating function (\ref{ap1}). It is useful to pass to the new variables
\begin{eqnarray} &&
\Psi=\frac{1}{\sqrt{\Delta}}\exp\left(
-\frac{\varphi}{4\nu}x^2\right)\tilde\Psi, \quad
P=\frac{1}{\sqrt{\Delta}}
\exp\left(\frac{\varphi}{4\nu}x^2\right)\tilde P,
\label{ha10} \end{eqnarray}
with the parameters that satisfy ordinary differential equations
\begin{equation}
\partial_t\Delta=\varphi\Delta, \quad 
\partial_t\varphi+\varphi^2-2\omega^3 c=0.
\label{ha11} \end{equation}
Since the quantities $\Delta$ and $\varphi$ are introduced 
by the differential equations where initial conditions are not fixed, 
then we can impose two arbitrary conditions without 
changing any observable quantity. Note that the equation (\ref{ha11})
for $\varphi$ describes the evolution of the linear 
profile $u=\varphi x$ of the velocity which the pumping $c$ 
would produce at $x\ll L$ in accordance with the inviscid equations. 

Let us introduce the dimensionless variables 
\begin{equation}
y=x/\Delta, \quad  
ds=-2\nu\Delta^{-2}dt.
\label{pp1} \end{equation} 
Then the system (\ref{ha8},\ref{ha9}) leads to the following 
equations for the functions $\tilde\Psi$ and $\tilde P$
\begin{eqnarray} &&
\partial_s\tilde\Psi+\frac{1}{2}\partial^2_y\tilde\Psi=0,
\label{ha13} \\ &&
\partial_s\tilde P-\frac{1}{2}\partial^2_y\tilde P
=-2\nu\delta(s)\partial_y\lambda\,\tilde\Psi^{-1}.
\label{ha12} \end{eqnarray}
Now, we see the essence of the transformation (\ref{ha10},\ref{ha11}): 
the equations (\ref{ha13},\ref{ha12}) are separated and do not contain  
the pumping explicitly. The expression (\ref{ap1}) is now rewritten as
\begin{equation}
Y=-\frac{i}{2\nu}\int dy\,
\tilde P(s,y)\tilde\Psi(s,y)\exp(-\tilde\eta y^2/2),
\label{app} \end{equation}
where $\tilde\eta=\Delta^2\eta$. The energy (\ref{hae}) 
in terms of the new variables takes the form
\begin{equation}
E=\frac{2i\nu}{\Delta^2}
\int dy\,\tilde P\partial_s\tilde\Psi
+i\varphi\int dy\,y\,\tilde P\,\partial_y\tilde\Psi
+\omega^3c^2-c\varphi^2.
\label{hae1} \end{equation}

The formal solution of the equations (\ref{ha13},\ref{ha12})
can be presented as
\begin{eqnarray} &&
\tilde\Psi(s,y)=\exp[-s\partial_y^2/2]\tilde\Psi(0,y),
\label{pp2} \\ &&
\tilde P(s,y)=-2\nu\exp[s\partial_y^2/2]
\tilde\Psi^{-1}(0,y)\partial_y\lambda.
\label{pp3} \end{eqnarray}
The expression (\ref{app}) becomes 
$Y=i\int dy\,\tilde\Psi(s,y)\exp(-\tilde\eta y^2/2)
\exp[s\partial_y^2/2]
\tilde\Psi^{-1}(0,y)\partial_y\lambda$.
Integrating over $dy$ by parts and substituting
(\ref{pp2}) we obtain 
\begin{eqnarray} &&
Y=i\int dy\,\tilde\Psi^{-1}(0,y)\partial_y\lambda
\exp\left\{-\frac{\tilde\eta}{2}
[y+s\partial_y]^2\right\}\tilde\Psi(0,y).
\label{pp7} \end{eqnarray} 
Here we used the following relation
\begin{eqnarray} &&
\exp(s\partial_y^2/2)\exp(-\tilde\eta y^2/2)
\exp(-s\partial_y^2/2)
=\exp\left\{-\frac{\tilde\eta}{2}
[y+s\partial_y]^2\right\}.
\label{pp6} \end{eqnarray}

At $t=0$ near the origin there is the shock (\ref{vm4}). Taking
into account also the linear profile we find at $x\ll L$
\begin{eqnarray} &&
\tilde\Psi(0,x)\propto\cosh\left(\frac{x}{x_0}\right)
\exp\left[\frac{1}{4\nu}(\varphi_0+\varphi_1)x^2\right],
\label{kk1}  \end{eqnarray}
where $\varphi_0$ is the value of $\varphi$ at $t=0$. We have introduced also 
the parameter $\varphi_1$ which accounts for the
deviation of the actual solution from $\tanh(x/x_0)$ at $t=0$,
$\varphi_1 x$ is the linear profile to be added to (\ref{vm4}). 
Starting from (\ref{kk1}) one can find the explicit expression 
for the generating function $Y$ via (\ref{pp7}). 
First we use the Hubbard-Stratonovich trick to convert
the second-order operator at the exponent into the first-order one:
\begin{eqnarray} &&
\exp\left(-\frac{\tilde\eta}{2}[y+s\partial_y]^2\right)=
\int\frac{d\xi}{\sqrt{2\pi\tilde\eta}}
\exp\left(-\frac{1}{2\tilde\eta}\xi^2+i\xi[y+s\partial_y]\right)
\nonumber \\ &&
=\int\frac{d\xi}{\sqrt{2\pi\tilde\eta}}
\exp\left(-\frac{1}{2\tilde\eta}\xi^2\right)
\exp\left(-\frac{1}{2}\xi^2s\right)
\exp\left(i\xi y\right)\exp\left(i\xi s\partial_y\right).
\label{pp8a} \end{eqnarray}
We used the relation $\exp(t_1y+t_2\partial_y)=\exp(t_1t_2/2)
\exp(t_1y)\exp(t_2\partial_y)$. Substituting (\ref{kk1},\ref{pp8a}) 
into (\ref{pp7}) we obtain after integrating over $\xi$
\begin{eqnarray} &&
Y=i\int dy\,\partial_y\lambda
\frac{1}{\sqrt{1+\tilde\eta s(1+\tilde\varphi_2s)}}
\exp\left[-\frac{\tilde\eta}{2}\frac{(1+\tilde\varphi_2s)^2y^2+s^2/y_0^2}
{1+\tilde\eta s(1+\tilde\varphi_2s)}\right]
\cosh\left[\frac{1}{1+\tilde\eta s(1+\tilde\varphi_2s)}\frac{y}{y_0}
\right]\cosh^{-1}\frac{y}{y_0},
\label{kk2} \end{eqnarray}
where $y_0=x_0/\Delta_0$ and 
$\tilde\varphi_2=\Delta_0^2(\varphi_0+\varphi_1)/(2\nu)$

\subsection{A solution}

To analyze the $t$-dependence of (\ref{kk2}) we should know
a solution of the system (\ref{ha11}) to restore $s(t)$ from 
(\ref{pp1}). For this, it is convenient to use another set of
dimensionless variables
\begin{equation}
\varphi=\frac{2\nu}{\Delta^2}\tilde\varphi
=\frac{2\omega}{\tilde\Delta^2}\tilde\varphi,\quad
a=\frac{2\nu}{\Delta^2}\tilde a
=\frac{2\omega}{\tilde\Delta^2}\tilde a,\quad
\Delta=L({\rm Re})^{-1/2}\tilde\Delta.
\label{ha14} \end{equation}
Then the equations (\ref{ha11}) are rewritten as
\begin{equation} 
\partial_s\tilde\Delta=-\tilde\varphi\tilde\Delta, \quad
\partial_s\tilde\varphi+\tilde\varphi^2
+\frac{1}{2}\tilde\Delta^6\tilde c=0, \quad
\omega c=\tilde\Delta^2\tilde c.
\label{ho36} \end{equation}
Let us write also the equation for the time $t$ following from (\ref{pp1})
\begin{equation}
\partial_st=-\frac{1}{2\omega}\tilde\Delta^2.
\label{ho41} \end{equation}
One derives from (\ref{ha10},\ref{ha13},\ref{kk1}) the following relations
\begin{equation}
\partial_s\ln\tilde\Psi(s,y=0)
=\frac{1}{2}(\tilde a-\tilde\varphi)
\Rightarrow \tilde a_0=-1/y_0^2-\tilde\varphi_1\;,
\label{ho3} \end{equation}
where $\tilde\varphi_1={\Delta^2}\varphi_1/2\nu$. 
After expanding (\ref{kk2}) we find 
\begin{eqnarray} &&
\tilde c=-\frac{i\nu}{2}
(1+\tilde\varphi_2 s)
\int dy\,\partial_y\lambda
\left[y^2(1+\tilde\varphi_2 s)+2s\frac{y}{y_0}\tanh
\frac{y}{y_0}\right],
\label{ha23} \end{eqnarray}
We see that $\tilde c$ is the function of the second order over $s$.
Substituting now the formal solution (\ref{pp2},\ref{pp3}) into (\ref{hae1}) 
we find after some calculations using the same tricks as above
\begin{equation}
\frac{E}{\omega}=\tilde\Delta^4\tilde c^2
-4\tilde c\tilde\varphi^2/\tilde\Delta^2
+4(\tilde\varphi/\tilde\Delta^2)\partial_s\tilde c
-(2/\tilde\Delta^2)\partial_s^2\tilde c.
\label{ha24} \end{equation}
One can directly check that the energy (\ref{ha24}) is conserved as a
consequence of (\ref{ho36},\ref{ho41}). Now, to establish the
$t$-dependence of $c$ we should solve the system (\ref{ho36})
and then reconstruct the function $s(t)$ from (\ref{ho41}).

We will see that the last term in the equation (\ref{ho36}) for
$\tilde\varphi$ can be neglected in the wide interval of changing 
$s$. Inside that interval (to be specified below),
the equation (\ref{ho36}) has a solution
\begin{equation}
\tilde\varphi=z^{-1},\quad\tilde\Delta= A_0z^{-1},
\quad z=s+\tilde\varphi_0^{-1},
\quad \omega t=-\frac{A_0^2}{2z}s\tilde\varphi_0,
\label{hq1} \end{equation}
implying that $\tilde\varphi_0<0$. We see from (\ref{hq1}) that $\tilde\Delta$ 
grows with decreasing $z$ and consequently the behavior (\ref{hq1}) is 
destroyed for small enough $z$. 

\subsection{Gradients}
\label{GRAD}

For the gradients, $\lambda$ is determined by (\ref{sh1}). Thus
one should substitute 
$\lambda={in}(2\nu\tilde a_0)^{-1}\delta'(y)$ into (\ref{kk2}):
\begin{eqnarray} &&
Y=\frac{n}{2\nu\tilde a_0}
\frac{(1+s\tilde\varphi_2)\tilde\eta}
{[1+s(1+s\tilde\varphi_2)\tilde\eta]^{3/2}}
\left\{\frac{1}{y_0^2}\frac{[2s+s^2(1+s\tilde\varphi_2)\tilde\eta]}
{[1+s(1+s\tilde\varphi_2)\tilde\eta]}+1+s\tilde\varphi_2\right\}
\exp\left[-\frac{1}{2y_0^2}\,\frac{s^2\tilde\eta}
{1+s\tilde\eta(1+\tilde\varphi_2s)}\right].
\label{kk3}\end{eqnarray}
Expanding (\ref{kk3}) one finds
\begin{eqnarray} &&
\tilde c=\frac{n}{2\tilde a_0}
(1+s\tilde\varphi_2)
\left[\frac{2s}{y_0^2}+(1+s\tilde\varphi_2)\right].
\label{pp9} \end{eqnarray}
It is time to recall that the energy should be equal to zero at our 
solution. At $t=0$ the value of $\tilde\Delta$ is very small and it 
is possible to neglect the first term in the right-hand side of (\ref{ha24}). 
Then, equating the residue to zero (and taking $s=0$) we find 
$\tilde\varphi_2\approx\tilde\varphi_0$ that is we can believe
$\tilde\varphi_1=0$. That means that with our accuracy it is
possible to neglect the linear profile in the the velocity at $t=0$
in comparison with the shock contribution. Neglecting $\tilde\varphi_1$
in (\ref{ho3}) we find also $\tilde a_0=-1/y_0^2$. Substituting the
relations into (\ref{pp9}) one obtains
\begin{equation}
\tilde c=n\tilde\varphi_0 z\left[
-s+\frac{1}{2\tilde a_0}\tilde\varphi_0z\right],
\label{ppa10} \end{equation}
where $z$ is determined by (\ref{hq1}). We see that $c$ tends to zero where 
$z\to 0$. It is related to growing $\tilde\Delta$ with decreasing $|z|$ 
since the first term in the expression (\ref{ha24}) for the energy $E$ 
cannot tend to infinity.

For $z\ll 1$ we can disregard the second term in (\ref{ppa10}) and write 
$\tilde c=-ns\tilde\varphi_0 z$. At $z\ll 1$, it follows from the definition
of $z$ that $s\approx-\tilde\varphi_0^{-1}$ and we conclude that 
$\tilde c\approx nz$. The behavior (\ref{hq1}) is observed if $|z|\gg z_1$ 
where $z_1\sim n^{1/3} A_0^2$ since at $z\sim -z_1$ the term 
proportional to $\tilde\Delta^6$ in (\ref{ho36}) is getting 
substantial. In the region $|z|\lesssim z_1$, both $\tilde\Delta$ and 
$\tilde\varphi$ do not vary essentially and can be estimated as follows 
\begin{equation}
\tilde\Delta\sim n^{-1/3} A_0^{-1}, \quad
\tilde\varphi\sim n^{-1/3} A_0^{-2}.
\label{ha31} \end{equation}
Using (\ref{ho36},\ref{hq1}) one obtains for $z_1\ll |z|\ll 1$
\begin{equation}
c=2nt.
\label{hq31} \end{equation}
This result is correct, in particular, for the time where
the width of $p$ is much larger than the width of the shock.
Then (\ref{hq31}) has to be compared with $c(t)=N^2f_3(t/T)$ what 
is the consequence of (\ref{vm3}) and of the definition (\ref{vm5}).
We thus conclude $N=n^{1/3}$. Then we find from (\ref{vm6})
\begin{equation} 
a(0)\sim\omega n^{2/3}{\rm Re}, \quad
T=-\omega n^{-1/3}.
\label{vm7} \end{equation}

It will be convenient for us to normalize $\Delta$ such that 
$\Delta=L$ at $z=0$, then from (\ref{ha14}) it 
follows that $\tilde\Delta=({\rm Re})^{1/2}$ there. 
Comparing this value with (\ref{ha31}) we conclude
\begin{equation}
A_0\sim{\rm Re}^{-1/2}n^{-1/3}
\quad z_1\sim n^{-1/3}{\rm Re}^{-1}.
\label{ha32} \end{equation}
Then we find from (\ref{ha14},\ref{hq1})
$\tilde a_0\sim\tilde\Delta_0^2{\rm Re}n^{2/3}\sim\tilde\varphi_0^2$.
It can be rewritten as $y_0\varphi_0\sim1$. Then the second contribution
to $\tilde c$ in (\ref{ppa10}) is $\sim n z^2$ and is consequently
much smaller than the first contribution at small $z$ what justifies
its neglecting. Now we can find the width of the field $p$ at $t=T$. We see 
that at $z\to 0$ all coefficients at $\tilde\eta$ in (\ref{kk3}) tends to zero 
excepting for the coefficient in the argument of the exponent which for small 
$z$ is $\exp[-s^2\tilde\eta/(2y_0^2)]$. Just the term determines the width of 
the field $p$ at small $z$, the width is $s\Delta/y_0\sim\Delta$ since at 
small $z$ $s\approx-\tilde\varphi_0^{-1}$ and $y_0\tilde\varphi_0\sim-1$. In
accordance with (\ref{ha14},\ref{hq1}) the width of the field $p$ increases 
with $|t|$: $\Delta\sim n^{1/3}L\omega t$ at $|t|\ll|T|$ and reaches the 
value $\Delta\sim L$ at $t\sim T$. 

Now we can estimate corrections associated with neglecting the first term
in the expression for the energy (\ref{ha24}) at $t=0$. Using (\ref{pp9})
we obtain from (\ref{ha24}) at $t=0$
$$ \frac{E}{\omega}=\frac{n^2}{4\tilde a_0^2}\tilde\Delta_0^4
-\frac{2n}{\tilde\Delta_0^2\tilde a_0}
\left(\frac{2}{y_0^2}\tilde\varphi_1+\tilde\varphi_1^2\right), $$
where $\tilde\varphi_1=\tilde\varphi_2-\tilde\varphi_0$ determines
the linear profile of the velocity. The term with $\tilde\varphi_1^2$ can 
be neglected here and we find $\tilde\varphi_1\sim nA_0^6\tilde\varphi_0^2$.
Let us estimate the correction to the expression (\ref{ppa10}) for $\tilde c$
following at nonzero $\tilde\varphi_1$ from (\ref{pp9}). 
First, due to the presence of $\tilde\varphi_1$,
$1/y_0^2$ differs from $-\tilde a_0$, see (\ref{ho3}). The ratio
$\tilde\varphi_1/a_0\sim A_0^6$ is negligible, see (\ref{ha32}).
Second, the term $1+s\tilde\varphi_2$ is now
$\tilde\varphi_0(z+\tilde\varphi_1\tilde\varphi_0^{-2})$.
The question is in the value of the correction at $z=z_1$, it is
estimated as $\tilde\varphi_1\tilde\varphi_0^{-2}/z_1\sim
{\rm Re}^{-2}n^{-2/3}$, see (\ref{ha32}), the correction is negligible. 
Thus we proved the correctness of the expression (\ref{ppa10}).
Note that the anomalously small value of the linear profile $\varphi_1$ 
shows that the formation of the shock is finished at $t=0$ since $\varphi_1$ 
directly determines the time derivative of the width of the shock:
$\partial_t\ln x_0=\varphi_1$. This is not surprising taking into account
extremal properties of the instanton.

\subsection{Velocity differences}

Let us apply the same scheme for the differences.
We introduce the dimensionless quantities
\begin{eqnarray} &&
\tilde w=\frac{\Delta_0w}{2\nu},\quad  
\tilde\rho=\frac{\rho}{\Delta_0}
\label{fdq1} \end{eqnarray}
so that we find from (\ref{sh2}) $\lambda={in}
[\delta(y+\tilde\rho)-\delta(y-\tilde\rho)]/({2\nu\tilde w})$.
Substituting this expression into (\ref{kk2}) we obtain
\begin{eqnarray} &&
Y=\frac{n}{\nu\tilde w}\tilde\eta(1+s\tilde\varphi_2)
\frac{[(1+s\tilde\varphi_2)\tilde\rho+s/y_0]}
{[1+s(1+s\tilde\varphi_2)\tilde\eta]^{3/2}}
\exp\left[-\frac{\tilde\eta[(1+s\tilde\varphi_2)\tilde\rho+s/y_0]^2}
{2[1+s(1+s\tilde\varphi_2)\tilde\eta]}\right].
\label{kk4} \end{eqnarray}
At calculating (\ref{kk4}) we substituted $\cosh$ by $\exp$
because of the inequality $\tilde\rho\gg y_0$ that means that
the separation $\rho$ between the points is much larger than
the (viscous) width of the front.

Expanding (\ref{kk4}) one finds
\begin{eqnarray} &&
\tilde c=\frac{n}{\tilde w}
(1+s\tilde\varphi_2)\left[
(1+s\tilde\varphi_2)\tilde\rho
+\frac{s}{y_0}\right].
\label{sv18} \end{eqnarray}
Again, using the relation $E=0$ with the energy (\ref{ha24}) we find 
that $\tilde\varphi_1=0$. Then disregarding $\tilde\varphi_1$ also in
(\ref{kk1}) we derive the relation $\tilde w=-2/y_0$. Finally,
\begin{equation}
\tilde c=\frac{n\tilde\varphi_0 z}{\tilde w}
\left(-2\tilde ws+\tilde\rho\tilde\varphi_0z\right).
\label{mm3} \end{equation}

Now, we should solve the equations (\ref{ho36}) with $\tilde c$
given by (\ref{mm3}). Again, we have the behavior (\ref{hq1})
stopped by the term with $\tilde c$ in (\ref{ho36}). To estimate
corresponding $z_1$ we should use the asymptotic of (\ref{mm3})
at small $z$. As we shall see, the leading term is the linear one
so that $\tilde c\sim n z$. Then we return to the same situation
as for gradients. That gives
\begin{equation}
w\sim L\omega n^{1/3} , 
\quad \tilde w\sim {\rm Re}^{1/2} A_0 n^{1/3},
\label{sv14} \end{equation}
where we used the condition $\rho\gg x_0$ to be satisfied.
To justify the above assertions we should check that at $z\sim -z_1$ 
the second term in (\ref{mm3}) is small in comparison with the first one. 
Utilizing $\tilde\rho\sim {\rm Re}^{1/2}{\rho}/(L A_0)$ and (\ref{ha32}) 
we obtain the estimates for the ratios of the noted terms
\begin{eqnarray} &&
\frac{\tilde\rho}{\tilde w}z_1
\sim\frac{\rho}{L A_0^2}\omega Tz_1
\sim\frac{\rho}{L}n^{-1/3}\ll 1,
\label{sv20}\end{eqnarray}
Thus the term is small since $\rho/L\ll1$ is implied.
All the other estimates are the same as for the case of gradients.

\subsection{Higher derivatives}

The above formalism allows us also to find the PDF tails for
the odd derivatives $u^{(k)}$ of the velocity. The average 
value $\langle[u^{(k)}]^n\rangle$ can be written like (\ref{si1},\ref{sio}) 
what leads to the saddle-point expression of the (\ref{vaa}) type. The 
corresponding instanton is determined by the equations (\ref{va2},\ref{vam}) 
where instead of (\ref{sh1},\ref{sh2}) one should take $\lambda$ to be
\begin{eqnarray} &&
\lambda=\frac{in}{u^{(k)}(0,0)}\delta^{(k)}(x).
\label{hd1}\end{eqnarray}
All above expressions containing $\lambda$ will be correct for odd $k$.
Substituting (\ref{hd1}) into (\ref{ha23}) one finds
\begin{eqnarray} &&
\tilde c=-\frac{k+1}{2}ns\tilde\varphi_0 z
\label{hd2}\end{eqnarray}
Performing the analysis similar to that of subsection \ref{GRAD}
we obtain that $N\sim [(k+1)n]^{1/3}$ and therefore the life time $T$
of the instanton is $T\sim\omega^{-1}[n(k+1)]^{-1/3}$. 
It becomes smaller for derivatives of higher order and therefore 
our approximation works better for larger $k$. Then from 
(\ref{vm4},\ref{vm6}) one finds the characteristic value 
$u^{(k)}(0,0)\sim N^{k+1}L^{1-k}\omega {\rm Re}^{k}$ leading to
\begin{eqnarray} &&
\left\langle[u^{(k)}]^n\right\rangle\sim\omega {\rm Re}^{k} 
L^{1-k} n^{(k+1)/3}.
\label{hd3}\end{eqnarray}
The result (\ref{hd3}) can be rewritten in terms of PDF: 
\begin{eqnarray} &&
{\cal P}\left(|u^{(k)}|\right)\propto
\exp\left[-C_k\left(\frac{|u^{(k)}|L^{k-1}}
{\omega{\rm Re}^k}\right)^{3/(k+1)}\right].
\label{hd4}\end{eqnarray}
Note that the non-Gaussianity increases with increasing $k$. On the other hand,
the higher $k$ the more distant is the validity region of (\ref{hd4}): 
$u^{(k)}\gg u^{(k)}_{\rm rms}\sim L^{1-k}\omega{\rm Re}^k$.

\subsection{Extremum action}

Let us now calculate the extremum value of the action (\ref{vae4}).
For $s\lesssim|\tilde\varphi_0|^{-1}$ one can rewrite it using 
(\ref{ha10},\ref{ha11})
\begin{equation}
i{\cal I}_{\rm extr}=-i\int ds\,dy\,\tilde P\partial_s\tilde\Psi
+\int ds\,\tilde c(2\tilde\varphi^2-\tilde c\tilde\Delta^6)
-i\int ds dy y\tilde P\partial_y\tilde\Psi\tilde\varphi. 
\label{sv21} \end{equation}
Using now the expression (\ref{hae1}) for the energy $E$ we can
rewrite (\ref{sv21}) as
\begin{equation}
i{\cal I}_{\rm extr}=-i\int ds\,
\left(-\frac{\tilde\Delta^2}{2\omega}E-
\frac{1}{2}\tilde c^2\tilde\Delta^6+\tilde a\right).
\label{sv22} \end{equation}
Let us recall that the energy $E$ in (\ref{sv22}) is $0$. Now we can 
estimate all the terms in (\ref{sv22}) utilizing the
asymptotics (\ref{hq1}) valid at $z\gtrsim z_1$ where 
$z_1\sim n^{1/3} A_0^2$. Then $\int ds\,\tilde\Delta^2\sim n^{-1/3}$.
Substituting (\ref{sv18}) we find $\int ds\,\tilde c^2\tilde\Delta^6\sim n$.
The last contribution in (\ref{sv22}) can be rewritten as
$$ \int ds\,\tilde a=\int dt a\approx-2\ln[\Psi(0,0)/\Psi(T,0)], $$
the last relation is the consequence of (\ref{ha1},\ref{ha7}). We see that 
only one contribution to $i{\cal I}_{\rm extr}$ is proportional to $n$, it is 
$\sim n$. Thus with our accuracy we can neglect the contribution.

It is natural that $\rho$-dependence of ${\cal P}(w)$ cannot be found in a 
saddle-point approximation; as a predexponent, it can be obtained only at the
next step by calculating the contribution of fluctuations around the instanton
solution. This is consistent with the known fact that the scaling exponent
is $n$-independent for $n>1$: $\langle w^n(\rho)\rangle\propto\rho$.
All said above is applicable also for high derivatives.

\subsection{Statistics of velocities.}

Let us now consider the statistics of velocity $u$.
Equations for fields $P$ and $\Psi$ are the same as for odd derivatives:
\begin{eqnarray} &&
\partial_t\Psi-\nu\partial_x^2\Psi+\nu F\Psi=0,
\label{vv7} \\ &&
\partial_t P+\nu\partial_x^2P-\nu FP
-{2\nu}\lambda'(x)\delta(t)\Psi^{-1}=0.
\label{v1} \end{eqnarray}
while $\lambda=-{in}\delta(x)/u(0,0)$ is not an odd function of $x$
so that $F$ is different now. 
Since $\int p\,dx$ is nonzero then $F$ is a linear (rather than quadratic)
function of $x$ for narrow $p$:
\begin{eqnarray}&&
F(x)=\frac{\chi(0)}{\nu^2}bx,\quad b=-\frac{i}{2}\int dx p(x)
\label{v2}\end{eqnarray}
This substantially simplifies the calculations. Substitution which separates 
variables
\begin{eqnarray}&&
\Psi=\frac{1}{\Delta}\exp(\varphi x)\tilde\Psi(t,y),\quad
P=\Delta\exp(-\varphi x)\tilde P(t,y),
\label{v3}\end{eqnarray}
includes $y=x+\rho$, $\Delta$ and $\varphi$ which satisfy the
following equations
\begin{eqnarray}&&
\partial_t\varphi+\frac{\chi(0)}{\nu} b=0,\quad 
\partial_t\rho-2\nu \varphi=0,\quad
\partial_t \Delta+\nu\varphi^2\Delta=0.
\label{v5}\end{eqnarray}
The equation for $\Delta$ is separated and since $\Delta$ does not
enter any observable we can forget about it.
The equations on $\tilde\Psi$ and $\tilde P$ are 
\begin{eqnarray}&&
\partial_t\tilde\Psi-\nu\partial^2_y\tilde\Psi=0,\quad
\partial_t\tilde P-\nu\partial^2_y\tilde P=
-\frac{2in\nu}{u_0}\delta'(y-\rho_0)\delta(t)
\tilde\Psi^{-1}
\label{v6}\end{eqnarray}
Formal solutions of (\ref{v6}) are
\begin{eqnarray}&&
\tilde P=\frac{2i\nu n}{u_0}\exp(-\nu t\partial_y^2)
\delta'(y-\rho_0)\tilde\Psi^{-1}(0,y), \quad
\tilde\Psi=\exp(\nu t\partial_y^2)\tilde\Psi(0,y)
\label{v8}\end{eqnarray}
Substituting (\ref{v8}) into $b={i}\int dy y\tilde\Psi\tilde P/{4\nu}$ 
we obtain
\begin{eqnarray} &&
b=-\frac{n}{2u_0}\int dy
\delta'(y-\rho_0)\tilde\Psi^{-1}(0,y)\exp(-\nu t\partial_y^2)y\tilde\Psi(t,y)
\nonumber \\ &&
=-\frac{n}{2u_0}\int dy\delta'(y-\rho_0)
\left[y-2\nu t\frac{\partial_y\tilde\Psi(0,y)}{\tilde\Psi(0,y)}\right]
\label{v9}\end{eqnarray}
The conservation of the momentum (\ref{va9}) gives
\begin{eqnarray} &&
iJ=\int dx P\partial_x\Psi=\int dy \tilde P\partial_y\tilde\Psi=
\frac{2i\nu n}{u_0}\int dy \delta'(y-\rho_0)
\frac{\partial_y\tilde\Psi(0,y)}{\tilde\Psi(0,y)}=
-\frac{in}{u_0}\int dx \delta'(x)\left[u(0,x)+\varphi\right]=0,
\label{v10}\end{eqnarray}
what leads to $\partial_xu(0,0)=0$. That naturally 
means that the maximum of $u$ reaches the point $x=0$ at $t=0$. 
We see that $J$ is the same integral which enters (\ref{v9}).
Therefore, $b$ is a constant (given by it's value at $t=-0$):
\begin{eqnarray}&&
b=\frac{n}{2u_0}\int dy
\delta(y-\rho_0)
=\frac{n}{2u(0,0)}
\label{v13}\end{eqnarray}
Similarly to the consideration at Sec. \ref{subsec:LT}, we can now estimate the
value of $u(0,0)$. Velocity stretches the field $p$ so that the width of $p$
reaches $L$ at $T\simeq L/u(0,0)$ while the velocity itself is produced by
the pumping during the same time: $u(0,0)\simeq \chi(0)bT=\chi(0)nT/2u(0,0)
\simeq n\chi(0)L/u(0,0)$. That gives $u(0,0)\simeq L\omega n^{1/3}$.
As well as in  the preceding subsection,  
we can neglect the term with ${\cal I}_{\rm extr}$ in (\ref{vaa})
and conclude that
$${\cal P}(u)\sim \exp[-D(u/L\omega)^3]\ .$$
Here $D$ is some numerical factor of order unity which is determined by
the evolution at $t\simeq T$ i.e. by the large-scale behavior of the pumping.

\section{Conclusion}

At smooth almost inviscid ramps, velocity differences and gradients are 
positive and linearly related $w(\rho)\equiv u(\rho)-u(-\rho)\approx 
2\partial_xu\rho$ so that the right tails of PDF have the same form
$${\cal P}(\partial_xu)\sim \exp[-(\partial_xu/\omega)^3]\,,
\quad{\cal P}(w)\sim \exp[-(w/2\rho\omega)^3]\ .$$
Those tails are universal i.e. they are determined by a single characteristics
of the pumping correlation function $\chi(r)$, namely, by it's second
derivative at zero $\omega=[-(1/2)\chi''(0)]^{1/3}$. 
Contrary, the left tails contain nonuniversal constant which 
depends on a large-scale behavior of the pumping. The left tails
come from shock fronts where $w^2\simeq -\nu\partial_xu$ so that cubic tail 
for velocity differences corresponds to semi-cubic tail for gradients:
$${\cal P}(\partial_xu)\sim\exp[-C(-\partial_xu/\omega{\rm Re})^{3/2}\,,\quad
{\cal P}(w)\sim \exp[-B(w/L\omega)^3]\ .$$
The formula for ${\cal P}(w)$ is valid for $w\gg u_{\rm rms}$ 
where $u_{\rm rms}\sim L\omega$. It is natural to assume that the 
probability is small for both $u(\rho)$ and $u(-\rho)$ being large 
simultaneously. Therefore, ${\cal P}(w)$ should coincide there with 
a single-point ${\cal P}(u)$. Indeed, we saw that at $u\gg u_{\rm rms}$
$${\cal P}(u)\sim \exp[-D(u/L\omega)^3]\ .$$
Here, $B,C$ and $D$ are (related) nonuniversal constants dependent upon
the behavior of $\chi(r)$ at $r\simeq L$. Note that the cubic tail for 
the single-point PDF has been obtained for decaying turbulence 
by a similar method employing the saddle-point approximation in the 
path integral with time as large parameter \cite{Avel}. 
Also, our formula for the derivatives
${\cal P}(|u^{(k)}|)\propto
\exp[-C_k(|u^{(k)}|L^{k-1}/{\omega{\rm Re}^k})^{3/(k+1)}]$
coincides with that of \cite{Avel} for the particular case of white (in space)
initial conditions. That, probably, means that white-in-time forcing
corresponds to white-in-space initial conditions. One should be cautious, 
however, comparing  the results for forced and decaying turbulence.

Another important restriction is our assumption on delta-correlated
pumping. If the pumping has a finite correlation time $\tau$ then our
results, strictly speaking, are valid for $u,w\ll L/\tau$.

\acknowledgements

We are grateful to M. Chertkov, D. Khmelnitskii and R. Kraichnan for useful 
discussions. This work was partially supported by the Rashi Foundation (G. F.), 
by the Minerva Center for Nonlinear Physics (I. K. and V. L.), 
and by the Minerva Einstein Center (V. L.).

\appendix

\section{}

Consistent account of fluctuations and calculation
of predexponent in PDF will be the subject of future work.
Yet one may worry if it is possible that fluctuations destroy instanton
i.e. give some divergent contribution into the action. That may happen due 
to some hidden symmetries and particular fluctuations that are deformations
along symmetry group. A special feature of our problem is
associated with $\delta$-correlated character of the pumping which 
leads to a quasi-symmetry of the gauge-invariance type. Namely, under
transformations
\begin{eqnarray} &&
p(t,x)\to p(t,x-r),\quad
u(t,x)\to u(t,x-r)+\partial_t{r}, \quad r=r(t),
\label{be9} \end{eqnarray}
the action ${\cal I}=\int dt\,{\cal L}$ determined by Lagrangian
(\ref{va1}) is transformed as
${\cal I}\to {\cal I}+\int dt\,dx\, p\,\partial_t^2 r$.
This additional term is equal to zero for functions $p$ satisfying
\begin{equation}
\partial_t^2\!\!\int dx\,p=0.
\label{be7} \end{equation}
For antisymmetric objects [gradients $u'$ and 
differences $w$ with $\lambda$ being $\delta'(x)$ and $\delta(x+\rho)-
\delta(x-\rho)$ respectively], $\lambda(x)=-\lambda(-x)$.
Then the integral $\int dt\,dx\,\lambda u$ entering (\ref{si1},\ref{sio})
is invariant under the transformation (\ref{be9}) if we require
$r(t=0)=0$ which will be implied below.

We thus have to do with the quasi-gauge transformation 
which left the effective action invariant on a wide 
subclass of functions. That means that the integration over the 
quasi-gauge degree of freedom should be performed exactly.
The conventional way to integrate over the gauge degree of freedom
is to pass to the action with a fixed gauge, excluding the volume of the 
corresponding gauge group. To perform the program for the transformation
(\ref{be9}) we include into the integrand in (\ref{si1},\ref{sio}) 
the additional factor
\begin{eqnarray} &&
\int{\cal D}r\,\delta[u(t,r)-\partial_t{r}] {\cal J}, \quad
{\cal J}=\det\left|\left|\frac{\delta [u(t,r)
-\partial_t{r}]}{\delta r}\right|\right|,
\label{ja2} \end{eqnarray}
where the integration is performed over functions $r(t)$. The factor 
(\ref{ja2}) is equal to unity by its construction and therefore does not 
influence ${\cal Z}$. Let us now change the order of integration to
$\int{\cal D}r\int{\cal D}u{\cal D}p$, and perform the transformation 
(\ref{be9}). Then the effective action ${\cal I}$ is shifted and 
$\delta[u(t,r)-\partial_t{r}]\to\delta[u(t,0)]$.
The Jacobian ${\cal J}$ introduced by (\ref{ja2}) is reduced to
\begin{equation}
{\cal J}\to\det[-\partial_t+\partial_xu(t,x)|_{x=0}].
\label{ja3}\end{equation}
Then the integration over $r$ gives the functional $\delta$-function
\begin{equation}
\int{\cal D}r\exp\left(i\int dt\,dx\,\partial_t^2r\,p\right)
\Rightarrow\delta\left[\int dx\,\partial_t^2 p(t,x)\right],
\label{ja7} \end{equation}
imposing the constraint on the field $p$.

Now, we should find the Jacobian (\ref{ja3}) which depends on the 
regularization of the transformation (\ref{be9}). Really, this 
regularization is fixed by the regularization of the integral 
(\ref{si1},\ref{sio}). The point is that at deriving (\ref{va1}) from 
the equation (\ref{be1}) the retarded regularization was implied (otherwise, 
some additional $u$-dependent term appears originating from the corresponding
Jacobian \cite{DP78}). The retarded regularization means that at 
discretizing time we should substitute
\begin{eqnarray} &&
\partial_tu+u\partial_xu
\Rightarrow \frac{u_n-u_{n-1}}{\epsilon}+u_{n-1} u'_{n-1},
\label{ja4}\end{eqnarray}
where $\epsilon$ is the step in time and $u'=\partial_xu$. 
The discretized version of the transformation (\ref{be9}) is
\begin{equation}
u_n(x)\to u_n(x-r_n)+\dot{r}_n,
\label{ja8} \end{equation}
where $\dot{r}=\partial_tr$ and $\dot{r}_n$ is the expression to be
determined. Substituting (\ref{ja8}) into the right-hand side of 
(\ref{ja4}) we obtain 
\begin{eqnarray} &&
\frac{u_n-u_{n-1}}{\epsilon}+u_{n-1}u'_{n-1}\to
\frac{1}{\epsilon}[u_n(x-r_n)-u_{n-1}(x-r_{n-1})]+
u_{n-1}u'_{n-1}+\dot{r}_{n-1}u'_{n-1}+\ddot{r}
\nonumber \\ &&
+\frac{1}{\epsilon}[u_n(x-r_n)-u_{n-1}(x-r_{n-1})]-
\frac{1}{\epsilon}(r_n-r_{n-1})u'_{n-1}
+\dot{r}_{n-1}u'_{n-1}+u_{n-1}u'_{n-1}+\ddot{r}.
\label{ja5} \end{eqnarray}
We see that the left-hand side of (\ref{ja5}) is preserved (up to terms 
$\ddot{r}$) if $\dot{r}_{n-1}=(r_n-r_{n-1})/{\epsilon}$ or
\begin{eqnarray}&&
\dot{r}_n=(r_{n+1}-r_n)/{\epsilon}.
\label{ja5a}\end{eqnarray}
Note that because of the condition $r(0)=0$ the contribution to
the Jacobian will be different depending on whether we consider positive 
or negative time. At $t<0$ with the regularization rule (\ref{ja5a}), 
Jacobian (\ref{ja3}) equals to
\begin{eqnarray}
{\cal J}=\det\left( \begin{array} {lcccccr}
\dots &\dots &\dots &\dots &\dots &\dots &\dots \\
\dots & 0 & 1/\epsilon+u'_n & -1/\epsilon & 0 & 0 & \dots \\
\dots & 0 & 0 & 1/\epsilon+u'_{n+1} & -1/\epsilon & 0 & \dots \\
\dots &\dots &\dots &\dots &\dots &\dots &\dots 
\end{array} \right)\Rightarrow 
\exp\left[\int dt \,\partial_x u(t,0)\right].
\label{ja10} \end{eqnarray}
At deriving (\ref{ja10}) we have not interested in factors which 
do not depend on the field $u$. All the factors should be absorbed into 
the normalization constant ${\cal Z}_0$ introduced in (\ref{si1},\ref{sio}). 
At $t>0$ we should go in the positive direction in time stating from
$r(0)=0$. Then we deal with the retarded regularization which does not
produce the contribution to the Jacobian (\ref{ja10}).
With the Jacobian, the action acquires the form
\begin{eqnarray} &&
{\cal I}\equiv \int dt\,{\cal L}, \quad {\cal L}=
\int dx\,(p\partial_t u+pu\partial_xu-\nu p\partial_x^2u)
\nonumber \\ &&
+\frac{i}{2}\int dx_1dx_2p_1\chi_{12}p_2-i\theta(-t)\partial_x u(t,0),
\label{va1'} \end{eqnarray}
where $\theta$ is the step function.

Now, at calculating the path integral (\ref{si1},\ref{sio}) one
should use only functions satisfying (\ref{be7}) and $u(t,0)=0$.
The instanton equation for $p$ will acquire the additional term
\begin{eqnarray} &&
\partial_t p+u\partial_xp
+\nu\partial_x^2p-i\theta(-t)\delta'(x)
=\delta(t)\lambda(x)
\nonumber \end{eqnarray}
which leads to a nontrivial vacuum:
\begin{eqnarray} &&
u\partial_xu-\nu\partial_x^2u
=-c\partial_x\chi(x),
\label{va6} \\ &&
u\partial_xp+\nu\partial_x^2p=i\delta'(x).
\label{va7} \end{eqnarray}
The requirements that $u(x)$ is a bounded function and $p(x)$ tends to zero 
where $x\to\infty$ (to provide the finite value for $c$), together with the 
gauge $u(t,0)=0$, determine the unique solution of (\ref{va6},\ref{va7}). 
The equation (\ref{va6}) has the first integral:
\begin{equation}
u^2/2-\nu(\partial_xu-a)+c[\chi(x)-\chi(0)]=0\ .
\label{va8} \end{equation}
If $x\to\pm\infty$ then $u\to\pm u_\infty=\pm\sqrt{2c\chi(0)-2\nu a}$.
The second term under the square root is a small correction of the order 
of ${\rm Re}^{-1}$. For small $x$, we have $u=ax$ and we find from (\ref{va8}) 
the relation $a^2=2c\omega^3$. Substituting $u\to ax$ into (\ref{va7}), 
multiplying it by $x$ and integrating over $x$, we find $2ac=1$. 
Then $c=1/(2\omega)$ and $a=\omega$, that means particularly
$u_\infty\sim L\omega$ which is $u_{rms}$. To obtain the form of $p(x)$, we
substitute $u\to ax$ into (\ref{va7}) and solve the equation:
\begin{equation} 
p=i\sqrt\frac{a}{2\pi\nu^3}{\rm sign}(x)
\int\limits_{|x|}^\infty d\xi\,\exp
\left(-\frac{a}{2\nu}\xi^2\right)\ .
\label{va10} \end{equation}
The width of $p(x)$ is $L/{\rm Re}$ i.e. it is small indeed. The energy of the
vacuum is finite $E=3\omega/4$.

It is important to realize that it is the ``gauge'' term in the right-hand side
of the equation for $p$ that makes vacuum non-trivial i.e. not given just by
$u=0$ and $p=0$. In our formalism, the breakdown of Galilean invariance, 
discussed by Polyakov \cite{Pol95}, manifests itself due to fixing gauge 
and making an explicit account of fluctuations of gauge degrees of freedom.

Now, we can repeat all the above instanton consideration with a new vacuum 
and the additional term in the equation for $p$. That gives no contribution
at the main order in large $n$.


\begin{references}

\bibitem{Burg} 
J.M. Burgers, {\it The Nonlinear Diffusion Equation} (Reidel, Dordrecht 1974).
\bibitem{Kra68} R. Kraichnan,
Phys. Fluids {\bf11}, 265 (1968).
\bibitem{GK93} T. Gotoh and R. Kraichnan,
Phys. Fluids {\bf 5}, 445 (1993).
\bibitem{Goto94} T. Gotoh,
Phys. Fluids {\bf 6}, 3985 (1994).
\bibitem{Pol95} 
A.M. Polyakov, Phys. Rev. E {\bf52}, 6183 (1995).
\bibitem{BMP95} J. Bouchaud, M. M\'ezard and G. Parisi,
Phys. Rev. E {\bf52}, 3656 (1995).
\bibitem{CY95} A. Chekhlov and V. Yakhot, Phys. Rev. E {\bf52}, 5681 (1995).
\bibitem{95KYC}  R.~Kraichnan, V.~Yakhot and S.~Chen, 
Phys.~Rev.~Lett. {\bf75}, 240 (1995).
\bibitem{95CFKL}
M. Chertkov, G. Falkovich, I. Kolokolov, V. Lebedev, 
Phys. Rev. E {\bf 52}, 4294 (1995).
\bibitem{95GK}
K. Gaw\c{e}dzki and A. Kupiainen, 
Phys. Rev. Lett. {\bf 75}, 3834 (1995)
\bibitem{95SS} 
B. Shraiman and E. Siggia, C.R. Acad. Sci. Paris, {\bf321}, 279 (1995).
\bibitem{95FKLM} 
G. Falkovich, I. Kolokolov, V. Lebedev and A. Migdal,
Phys. Rev. E (submitted), chao-dyn/9512006.
\bibitem{95GM} 
V. Gurarie and A. Migdal, Phys. Rev. E (submitted), chao-dyn/9512012.
\bibitem{GK96} 
T. Gotoh and R. Kraichnan, (in preparation)
\bibitem{73MSR}
P.~C. Martin, E.~D. Siggia, and H.~A. Rose, Phys.~Rev.~A {\bf 8}, 423 (1973).
\bibitem{76Dom}
C.~de~Dominicis, J.~Physique (Paris) {\bf 37}, c01-247 (1976).
\bibitem{76Jan}
H. Janssen, Z.~Phys.~B {\bf 23}, 377 (1976).
\bibitem{DP78}
C. de Dominicis and L. Peliti,
Phys. Rev. B {\bf 18}, 353    (1978).
\bibitem{Avel} M. Avellaneda, R. Ryan and E. Weinan,
Phys. Fluids {\bf7}, 3067 (1995).

\end{references}
\end{document}